\documentstyle[11pt,epsf]{article}
\def\beq{\begin{eqnarray}}
\def\eeq{\end{eqnarray}}
\def\bea{\begin{eqnarray*}}
\def\eea{\end{eqnarray*}}

\def\centeron#1#2{{\setbox0=\hbox{#1}\setbox1=\hbox{#2}\ifdim
\wd1>\wd0\kern.5\wd1\kern-.5\wd0\fi
\copy0\kern-.5\wd0\kern-.5\wd1\copy1\ifdim\wd0>\wd1
\kern.5\wd0\kern-.5\wd1\fi}}
\def\ltap{\;\centeron{\raise.35ex\hbox{$<$}}{\lower.65ex\hbox{$\sim$}}\;}
\def\gtap{\;\centeron{\raise.35ex\hbox{$>$}}{\lower.65ex\hbox{$\sim$}}\;}

\def\singleandthirdspaced{\baselineskip=\normalbaselineskip\multiply
    \baselineskip by 130\divide\baselineskip by 100}

\def\dslash{\not{\hbox{\kern-2pt $\partial$}}}
\def\Dslash{\not{\hbox{\kern-4pt $D$}}}
\def\Oslash{\not{\hbox{\kern-4pt $O$}}}
\def\Qslash{\not{\hbox{\kern-4pt $Q$}}}
\def\pslash{\not{\hbox{\kern-2.3pt $p$}}}
\def\kslash{\not{\hbox{\kern-2.3pt $k$}}}
\def\qslash{\not{\hbox{\kern-2.3pt $q$}}}

\newcommand{\newc}{\newcommand}
\newc{\qbar}{{\overline q}}
\newc{\Kahler}{K\"ahler }
\newc{\deltaGS}{\delta_{\rm GS}}
\begin{document}
\begin{titlepage}
\begin{flushright}
{\large hep-th/yymmnnn \\ SCIPP-2006/13\\
}
\end{flushright}

\vskip 1.2cm

\begin{center}

{\LARGE\bf Locality and the classical limit of quantum systems}

\vskip 1.4cm

{\large Tom Banks}
\\
\vskip 0.4cm

 {\it Department of Physics and SCIPP\\
 University of California, Santa Cruz, CA 95064\\
 E-mail: {banks@scipp.ucsc.edu}\\
 {\it and}\\
 Department of Physics and NHETC, Rutgers University\\
 Piscataway, NJ 08540 }

 \vskip 4pt

\vskip 1.5cm

\begin{abstract}
 I argue that conventional estimates of the criterion for classical
 behavior of a macroscopic body are incorrect in most circumstances,because they do not
 take into account the locality of interactions, which characterizes the
 behavior of all systems described approximately by local quantum
 field theory.  The deviations from classical behavior of a macroscopic body,
 except for those that can be described as classical uncertainties
 in the initial values of macroscopic variables,are {\it exponentially}
 small as a function of the volume of the macro-system in microscopic units.
 Conventional estimates are correct only when the internal degrees
 of freedom of the macrosystem are in their ground state, and the classical
 motion of collective coordinates is adiabatic. Otherwise, the
 system acts as its own environment and washes out quantum phase
 correlations between different classical states of its collective
 coordinates. I suggest that it is likely that we can only achieve
 meso-scopic superpositions, for systems which have topological
 variables, and for which we can couple to those variables without
 exciting phonons.

\end{abstract}

\end{center}

\vskip 1.0 cm

\end{titlepage}
\setcounter{footnote}{0} \setcounter{page}{2}
\setcounter{section}{0} \setcounter{subsection}{0}
\setcounter{subsubsection}{0}

\singleandthirdspaced

\section{Classical behavior in the non-relativistic quantum mechanics
of particles}

In standard texts on non-relativistic quantum mechanics the
classical limit is described via examples and via the WKB
approximation. In particular, one often describes the spreading of
the wave packet of a free particle, and estimates it as a function
of time and the particle mass $M$. There is nothing wrong with the
mathematics done in these texts, but the implication that these
estimates provide the basis for an understanding of why classical
mechanics is such a good approximation for macroscopic objects is
not correct and therefore misleading. In particular it leads one to
conclude that the corrections to decoherence for a wave function
describing a superposition of two different macroscopic states is
power law in the mass. I would aver that this mistake forms part of
the psychological unease that many physicists feel about the
resolution of Schr\"{o}dinger's cat paradox in terms of the concept
of decoherence.

These estimates have also led to recent experimental proposals to
demonstrate quantum superposition of states of variables which are
\lq\lq almost macroscopic".  I will argue that no such demonstration
is possible, without extreme care taken to keep the constitutents of
the macrosystem in their microscopic ground state. The essence of my
argument is that the essential variable that controls the approach
to the classical limit, is the number of localizable constituents of
large quantum system. In a macroscopic material this would be
something like the number, $N$, of correlation volumes contained in
the sample.  Away from critical points, the correlation volume is
microscopically small, and we are roughly counting the number of
atoms.

Indeed, all previous discussions also identify this number as the
crucial parameter. These discussions identify a variety of
collective classical variables, like the center of mass of the
system, and note that the effective Lagrangian for these variables
has a factor of $N$ in it.  For the center of mass, this is simply
the statement that the mass is large.  The traditional argument
simply studies the quantum mechanics of these collective variables
and estimates the corrections to classical predictions, which are
typically power law in the large, extensive parameters.  Estimates
based on these ideas have led to the suggestion that plausible
extensions of current experiments can reach the limit of quantum
coherence for collective coordinates of systems with dimensions of
millimeters.  The failure to observe such correlations might be
taken to mean that there is some fundamental error in applying
quantum mechanics to macroscopic systems, as has been proposed by
Penrose, Leggett and others.

The essential point of this paper is that ``small" corrections to
this collective coordinate approximation completely invalidate this
argument. It is not that the classical dynamics is not a good
approximation to the quantum mechanics of the collective variables.
What is not a good approximation is to neglect the back reaction of
the collective variables on the huge set of other degrees of freedom
in the macroscopic object.  Locality ensures that external forces
acting on the macroscopic body affect the collective coordinate
through collective interactions with individual constituents, which
then give rise to terms in the Hamiltonian coupling the collective
coordinate to the constituents.  In effect, different classical
motions of the collective coordinate give rise to different, time
dependent, Hamiltonians for the constituents.  These extra terms are
small, inversely proportional to powers of extensive parameters.

However, typical macro-systems have a finite microscopic correlation
length.  The wave function of the system is a sum of terms which are
products of individual cluster wave functions for a localized
microscopic subset of the constituents.  This idea is the basis for
approximate variational calculations like the Hartree-Fock or
Jastrow approximations. As a consequence of the small corrections
described in the previous paragraph, the individual cluster wave
functions will be modified by a small amount and the overlap between
wave functions for two different classical trajectories of the
collective coordinate will be proportional to $1 - \epsilon$ where
$\epsilon$ is a measure of the strength of the perturbation that
leads to non-uniform motion of the center of mass. However, because
the full many body wave function is a product of $o(N)$ cluster wave
functions, the overlap is of order $(1 - \epsilon)^N$, {\it which is
exponentially small in the volume of the system measured in
microscopic units}. In other words, {\it for a macroscopic body,
different classical trajectories of a collective coordinate divide
the system into different approximate super-selection sectors.}

One can argue, using the methods of quantum field theory and
statistical mechanics, that the time that it takes to observe phase
correlations between different approximate superselection sectors is
of order $10^{ c N}$ where $c$ is a constant of order one. This is
true as long as one is in a regime where the density of states of
the microscopic degrees of freedom is large, {\it i.e.} that the
state of the system is a superposition of a densely spaced set of
eigenstates, which behaves in a manner describable by statistical
mechanics. Note that the ratio between the current age of the
universe and the Planck time is a mere $10^{61}$, so that even for a
moderately large system containing $N \sim 10^3$ correlation
volumes, this time is {\it so long that it is essentially the same
number of Planck times as it is ages of the universe. No imaginable
experiment can ever distinguish the quantum correlations between
different states of the collective coordinates of a macro-system.}
The extraordinary smallness of such double exponentials defeats all
of our ordinary intuitions about ordinary physics.  Over such long
time scales, many counter-intuitive things could happen. For
example, in a hypothetical classical model of a living organism made
of this many constituents, or in a correct quantum model, the
phenomenon of Poincare recurrences assures that given (exponentially
roughly) this much time,the organism could spontaneously self
assemble,out of a generic initial state of its constituents. So much
for Schr\"{o}dinger's cat.

Another way of phrasing the same arguments comes from the vast
literature on {\it decoherence}, which also introduces an important
concept I have not yet emphasized. This is the fact that an
approximate superselection sector is not a single state, but
actually a vast ensemble of order $10^{cN}$ states, which share the
same value of the collective coordinate.  In the decoherence
literature, it is argued that rapid changes in the micro-state of a
macroscopic environment wipe out the quantum phase correlations
between {\it e.g.} states with two different positions of a
macroscopic pointer, which have been put into a {\it
Schr\"{o}dinger's cat} superposition via interaction with some
micro-system.  Another way to state the conclusions of the previous
paragraph is simply to say that the constituents of a macroscopic
body serve as an environment, which serves to decohere the quantum
correlations between the macro-states of collective coordinates.
Unlike typical environments, which one might hope to eliminate by
enclosing the system in a sufficiently good vacuum, the inherent
environment of a macro-system cannot be escaped. The collective
variables exist and behave as they do, because of the properties of
the environment in which they are embedded.  It is only when the
macroscopic system is held in its ground state, during experiments
in which the dynamics of the collective variables is probed, that
conventional estimates of quantum coherence for the collective
coordinate wave function are valid.

In this introductory section, I will fill in the argument that
conventional estimates of quantum corrections to classical behavior
are wrong, using standard ideas of non-relativistic quantum
mechanics. In the remainder of the paper I will discuss the basis
for these calculations in quantum field theory. This will also
remove the necessity to resort to Hartree-Fock like approximations
to prove the point directly in the non-relativistic formalism. As
noted, the essential point of the argument is that we must take into
account the fact that a macroscopic object is made out of a huge
number, which generally I will take to be $
> 10^{20}$, of microscopic constituents, in order to truly understand
its classical behavior. I will argue that, as a consequence, the
overlaps between states where the object follows two macroscopically
different trajectories, as well as the matrix elements of all local
operators\footnote{In this context local means an operator which is
a sum of terms, each of which operates only on a few of the
constituent particles. A more precise, field theoretic, description
will be given in the next section.} between such states, are of
order
$$e^{ - 10^{20}} .$$

Consider then, the wave function of such a composite of $N \gg 1$
particles, assuming a Hamiltonian of the form
$$H = \sum {{\overrightarrow{p}_i^2}\over 2 m_i} + \sum V_{ij} (x_i -
x_j).$$ Apart from electromagnetic and gravitational forces, the two
body potentials are assumed to be short ranged. We could also add
multi-body potentials, as long as the number of particles that
interact is $\ll N$\footnote{Or that the strength of $k$ body
interactions fall off sufficiently rapidly with $k$ for $k > N_0 \ll
N$.}.

The Hamiltonian is Galilean invariant and we can separate it into
the kinetic energy of the center of mass, and the Hamiltonian for
the body at rest. The wave function is of the form

$$\psi (X_{cm}) \Psi (x_i - x_j) .$$ $\Psi$ is a general function of coordinate
differences.  In writing the Schrodinger equation we must choose $N
- 1$ of the coordinate differences as independent variables.  If the
particles are identical, this choice obscures the $S_N$ permutation
symmetry of the Hamiltonian. One must still impose Bose or Fermi
statistics on the wave functions.  This is a practical difficulty,
but not one of principal. We now want to compare this wave function
with the internal wave function of the system when the particle is
not following a straight, constant velocity trajectory. In order to
do this, we introduce an external potential $U(x_i)$. It is {\it
extremely} important that $U$ is not simply a function of the center
of mass coordinate but a sum of terms denoting the interaction of
the potential with each of the constituents. This very natural
assumption is derivable from local field theory: the external
potential must interact locally with ``the field that creates a
particle at a point".  So we assume
$$U = \sum u_i(x_i),$$ where we have allowed for the possibility,
{\it e.g.} that the external field is electrical and different
constituents have different charge.

To solve the external potential problem, we write $x_i = X_{cm} +
\Delta_i$ and expand the individual potentials around the center of
mass, treating the remaining terms as a small perturbation. We then
obtain a Hamiltonian for the center of mass, which has a mass of
order $N$, as well as a potential of order $N$.  The large $N$ limit
is then the WKB limit for the center of mass motion. The residual
Hamiltonian for the internal wave function has small external
potential terms, whose coefficients depend on the center of mass
coordinate.

The Schrodinger equation for the center of mass motion thus has
solutions which are wave functions concentrated around a classical
trajectory $X_{cm} (t)$ of the center of mass, moving in the
potential $\sum u_i (X_{cm}) $\footnote{See \cite{nauenberg} for a
construction of such wave functions for the Coulomb/Newton
potential.}. These wave functions will spread with time in a way
that depends on this potential. For example, initial Gaussian wave
packets for a free particle will have a width, which behaves like
$\sqrt{t/N m}$ for large $t$, where $m$ is a microscopic mass scale.
The fact that this is only significant when $t \sim N$ is the
conventional explanation for the classical behavior of the center of
mass variable.

In fact, this argument misses the crucial point, namely that the
small perturbation, which gives the Hamiltonian of the internal
structure a time dependence, through the appearance of $X_{cm} (t)$,
is not at all negligible. To illustrate this let us imagine that the
wave function at rest has the Hartree-Fock form, an anti-symmetrized
product of one body wave functions $\psi_i (x_i)$, and let us
characterize the external potential by a strength $\epsilon$. In the
presence of the perturbation, each one body wave function will be
perturbed, and its overlap with the original one body wave function
will be less than one. {\it As a consequence, the overlap between
the perturbed and unperturbed multi-body wave functions will be of
order $(1 - \epsilon)^N$}.  This has the exponential suppression we
claimed, as long as $\epsilon \gg {1\over N}$. It is easy to see
that a similar suppression obtains for matrix elements of few body
operators. One can argue that a similar suppression is obtained for
generalized Jastrow wave functions, with only few body correlations,
but a more general and convincing argument based on quantum field
theory will be given in the next section. Here we will follow
through the consequences of this exponential suppression.

The effect is to break up the full Hilbert space of the composite
object in the external potential, into {\it approximate
super-selection sectors} labeled by macroscopically different
classical trajectories $X_{cm} (t)$ (microscopically different
trajectories correspond to $\epsilon \sim {1\over N}$). We will
argue that local measurements cannot detect interference effects
between states in different super-selection sectors on times scales
shorter than $e^{10^{20}} $ (we leave off the obviously irrelevant
unit of time). That is to say, for all {\it in principle purposes},
a superposition of states corresponding to different classical
trajectories behaves like a classical probability distribution for
classical trajectories. The difference of course is that in
classical statistical physics one avers that {\it in principle} one
could measure the initial conditions precisely, whereas in quantum
mechanics the uncertainty is intrinsic to the formalism.

The argument for the exponentially large time scale has two parts,
each of which will be given in more detail below. First we argue
that it takes a time of order $N$, for a local Hamiltonian to
generate an overlap of order $1$ between two different
superselection sectors. Then we argue that most macroscopic objects
are not in their quantum ground state.  The typical number of
eigen-states present in the initial state of the object, or that can
be excited by the coupling to the time dependent motion of the
collective coordinate is of order $e^{c N}$. These states are highly
degenerate.  The time dependent Hamiltonian induced by the coupling
to the collective coordinate will induce a time dependent unitary
evolution on this large space of states, with a time scale of order
$1$ (in powers of $N$). Thus, there is a rapid loss of phase
coherence between the two super-selection sectors, while the
Hamiltonian is generating a non-trivial overlap between them. We
would have to wait for the motion on the Hilbert space to have a
recurrence before we could hope to see coherent quantum interference
between two states with different macroscopic motions of the
collective coordinate.  The shortest recurrence time is of order
$e^{cN}$.

Two paragraphs ago, I used the phrase {\it in principle} in two
different ways. The first use was ironic; the natural phrase that
comes to mind is {\it for all practical purposes}. I replace {\it in
practice} by {\it in principle} in order to emphasize that any
conceivable experiment that could distinguish between the classical
probability distribution and the quantum predictions would have to
keep the system isolated over times inconceivably longer than the
age of the universe. In other words, it is meaningless for a {\it
physicist} to consider the two calculations different from each
other. In yet another set of words; the phrase ``With enough effort,
one can in principle measure the quantum correlations in a
superposition of macroscopically different states", has the same
status as the phrase ``If wishes were horses then beggars would
ride".

The second use of {\it in principle} was the conventional
philosophical one: the mathematical formalism of classical
statistical mechanics contemplates arbitrarily precise measurements,
on which we superimpose a probability distribution which we
interpret to be a measure of our ignorance. In fact, even in
classical mechanics for a system whose entropy is order $10^{20}$,
this is arrant nonsense. The measurement of the precise state of
such a system would again take inconceivably longer than the age of
the universe.

This comparison is useful because it emphasizes the fact that the
tiny matrix elements between super-selection sectors are due to an
entropic effect. They are small because a change in the trajectory
of the center of mass changes the state of a huge number of degrees
of freedom. Indeed, in a very rough manner, one can say that the
time necessary to see quantum interference effects between two
macroscopically different states is of order the Heisenberg
recurrence time of the system. This is very rough, because there is
no argument that the order $1$ factors in the exponent are the same,
so the actual numbers could be vastly different. The important point
is that for truly macroscopic systems both times are
super-exponentially longer than the age of the universe.

The center of mass is one of a large number of {\it collective} or
{\it thermodynamic} observables of a typical macroscopic system
found in the laboratory. The number of such variables is a measure
of the number of macroscopic moving parts of the system. As we will
see, a system with a goodly supply of such moving parts is a good
measuring device. Indeed, the application of the foregoing remarks
to the quantum measurement problem is immediate. As von Neumann
first remarked, there is absolutely no problem in arranging a
unitary transformation which maps the state
$$\alpha |\uparrow \rangle > + \beta |\downarrow \rangle > \otimes |
Ready >,$$ of a microsystem uncorrelated with the $| Ready \rangle $
state of a measuring apparatus, into the correlated state
$$\alpha |\uparrow \rangle > \otimes | + \rangle + \beta |\downarrow \rangle > \otimes |
- >,$$ where $| +/- >$ are {\it pointer states} of the measuring
apparatus. If we simply assume, in accordance with experience, that
the labels $+/-$ characterize the value of a macroscopic observable
in the sense described above, then we can immediately come to the
following conclusions

\begin{itemize}

\item 1.The quantum interference between the two pieces of the wave
function cannot be measured on time scales shorter than the
super-exponential times described above. The predictions of quantum
mechanics for this state are identical {\it in principle} (first
usage) to the predictions of a classical theory that tells us only
the probabilities of the machine reading $+$ or $-$. {\it Like any
such probabilistic theory} the algorithm for interpreting its
predictions is to condition the future predictions on any actual
measurements made at intermediate times. This is the famous
``collapse of the wave function", on which so much fatuous prose has
been expended. It no more violates conservation of probability than
does throwing out those weather simulations, which predicted that
Hurricane Katrina would hit Galveston.

\item 2.One may worry that there is a violation of unitarity in this
description, because if I apply the {\it same} unitary
transformation to the states $|\uparrow \rangle \otimes | Ready
\rangle $ and $|\downarrow \rangle \otimes | Ready \rangle $,
individually, then I get a pair of states whose overlap is not
small. This seems like a violation of the superposition principle,
but this mathematical exercise has nothing to do with physics, for
at least two reasons. First the macro-states labeled by $+/-$ are
not single states, but huge ensembles, with $e^N$ members. The
typical member of any of these ensembles is a time dependent state
with the property that time averages of all reasonable observables
over a short relaxation time are identical to those in another
member of the ensemble. The chances of starting with the identical
$| Ready \rangle$ state or ending with the same $ | +/- \rangle$
states in two experiments with different initial micro-states, is
$e^{-N}$. Furthermore, and perhaps more importantly, the
experimenter who designs equipment to amplify microscopic signals
into macroscopic pointer readings, {\it does not} control the
microscopic interaction between the atoms in the measuring device
and {\it e.g.} the electron whose spin is being measured. Thus, in
effect, every time we do a new measurement, whether with the same
input micro-state or a different one, it is virtually certain that
the unitary transformation that is actually performed on the system
is a different one.

\end{itemize}

For me, these considerations resolve all the {\it angst} associated
with the Schr\"{o}dinger's cat paradox. Figurative superpositions of
live and dead cats occur every day, whenever a macroscopic event is
triggered by a micro-event. We see nothing remarkable about them
because quantum mechanics makes no remarkable predictions about
them. It never says ``the cat is both alive and dead", but rather,
``I can't predict whether the cat is alive or dead, only the
probability that you will find different cats alive or dead if you
do the same experiment over and over". Wave function collapse and
the associated claims of instantaneous action at a distance are
really nothing but the the familiar classical procedure of
discarding those parts of a probabilistic prediction, which are
disproved by actual experiments. This is usually called the use of
conditional probabilities, and no intellectual discomfort is
attached to it.

It is important to point out here that I am not claiming that any
classical probability theory could reproduce the results predicted
by quantum mechanics.  John Bell showed us long ago that this is
impossible, as long as we insist that our classical theory obey the
usual rules of locality.  My claim instead is that the correct
philosophical attitude toward collapse of the wave function is
identical to that which we invoke for any theory of probability.  In
either case we have a theory that only predicts the chances for
different events to happen, and we must continuously discard those
parts of the probability distribution, which predicted things that
did not occur.  The considerations of this paper show that when we
discard the dead cat part of the wave function after seeing that the
cat is alive, we are making mistakes about future predictions of the
theory that are in principle unmeasurable.

We are left with the discomfort Einstein expressed in his famous
aphorism about mythical beings rolling dice. Those of us who
routinely think about the application of quantum mechanics to the
entire universe, as in the apparently successful inflationary
prediction of the nature of Cosmic Microwave Background temperature
fluctuations, cannot even find comfort in the frequentist's fairy
tale about defining probability ``objectively" by doing an infinite
number of experiments. Probability is a guess, a bet about the
future. What is it doing in the most precisely defined of sciences?
I will leave this question for each of my readers to ponder in
solitude. I certainly don't know the answer.

Finally, I want to return to the spread of the wave packet for the
center of mass, and what it means from the point of view presented
here. It is clear that the uncertainties described by this wave
function can all be attributed to the inevitable quantum
uncertainties in the initial conditions for the position and
velocity of this variable. Quantum mechanics prevents us from
isolating the initial phase space point with absolute precision.
These can simply be viewed as microscopic initial uncertainties in
the classical trajectory $X_{cm} (t)$. In the WKB approximation, the
marginal probability distributions for position and momentum are
Gaussian, and there is a unique Gaussian phase space distribution
that has the same marginal probabilities.

If we wait long enough these uncertainties would, from a purely
classical point of view, lead to macroscopic deviations of the
position from that predicted by the classical trajectory we have
expanded around. The correct interpretation of this is that our
approximation breaks down over such long time scales. A better
approximation would be to decide that after a time long enough for
an initial microscopic deviation to evolve into a macroscopic one,
we must redefine our super-selection sectors. After this time,
matrix elements between classical trajectories that were originally
part of the same super-selection sector, become so small that we
much declare that they are different sectors.

Thus instead of, in another famous Einsteinian phrase, complaining
that the moon is predicted to disappear when we don't look at it
(over a time scale power law in its mass), we say that quantum
mechanics predicts that our best measurement of the initial position
and velocity of the moon is imprecise. The initial uncertainties are
small, but grow with time, to the extent that we cannot predict
exactly where the moon is. Quantum mechanics {\it does} predict,
that the moon has  (to an exponentially good approximation) followed
some classical trajectory, but does not allow us to say which one, a
long time after an initial measurement of the position and velocity.

Of course, if the constituents of the macroscopic body are kept in
their ground state during the motion, then we must treat the wave
function of the center of mass with proper quantum mechanical
respect, and the predictions of quantum interference between
different classical trajectories should be verifiable by experiment.
This is clearly impossible for the moon.  In a later section, I will
discuss whether it is likely to be true for mesoscopic systems
realizable in the laboratory.

\subsection{Bullets over Broad-slit-way }

To make these general arguments more concrete, let's consider
Feynman's famous discussion of shooting bullets randomly through a
pair of slits broad enough to let the bullets pass through. The
bullet moves in the $x$ direction, and we assume the initial wave
function of the center of mass of the bullet is spread uniformly
over the y coordinate distance between the slits. Then subsequent to
the passage through the slits, the wave function of the center of
mass is, to a good approximation, a superposition of two Gaussian
wave functions, centered around the two slit positions.  A
conventional discussion of this situation would solve the free
particle Schrodinger equation for this initial wave function and
compare the quantum mechanical probability distribution a later
times, with a classical distribution obtained by solving the
Liouville equation for a free particle, with initial position and
momentum uncertainties given by some positive phase space
probability distribution whose marginal position and momentum
distributions coincide with the squares of the position and momentum
space wave functions.

In the latter calculation, the term in the initial probability
distribution coming from the overlap of the Gaussians centered at
the two different slits is of order $e^{- (\frac {L}{w}})^2$, where
$L$ is the distance between the slits and $w$ their width. Liouville
evolution can lead to uncertainty about which slit the particle went
through in a time of order $\frac{2M L}{\hbar} $, just as in the
quantum calculation.   However, it gives rise to a different spatial
distribution of probability density, with no interference peaks.
Thus, for such times, the interference terms and exact Schrodinger
evolution give a different result from classical expectations with
uncertain initial conditions.

Now, let us take into account the fact that the two branches of the
center of mass wave function must be multiplied by wave functions of
the internal coordinates, which are in different super-selection
sectors.  As long as the micro-state is a superposition of internal
eigenstates coming from a band with exponentially large density of
states, it would be highly unnatural to assume that the micro-state
in the top slit is simply the space translation of that in the
bottom slit.  The probability for this coincidence is $e^{- c N}$.
It then follows from our previous discussion that we will have to
wait of order a recurrence time in order to have a hope that the
interference term in the square of the Schrodinger wave function is
not exponentially small.  The difference between quantum evolution
of the center of mass wave function, and classical evolution with
uncertain initial conditions is completely unobservable, except
perhaps at selected instants over super-exponentially long time
scales.

\section{\bf Quantum field theory}

I will describe the considerations of this section in the language
of relativistic quantum field theory. {\it A fortiori} they apply to
the non-relativistic limit, which we discussed in first quantization
in the previous section. They also apply to cutoff field theories,
with some kind of spatial cutoff, like a space lattice. The key
property of all these systems is that the degrees of freedom are
labeled by points in a fixed spatial geometry, with a finite number
of canonical bosonic or fermionic variables per point. The
Hamiltonian of these degrees of freedom is a sum of terms, each of
which only couples together the points within a finite
radius\footnote{Various kinds of exponentially rapid falloff are
allowed, and would not effect the qualitative nature of our
results.} In the relativistic case of course the Hamiltonian is an
integral of a strictly local Hamiltonian density.

Let us first discuss the ground state of such a system. If the
theory has a mass gap, then the ground state expectation values of
products of local operators fall off exponentially beyond some
correlation length $L_c$. If $d$ is the spatial dimension of the
system,and $V$ is a volume $\gg L_c^d$, define the state
$$|\phi_c , V \rangle ,$$ as the normalized state with minimum
expectation value of the Hamiltonian, subject to the constraint that
$$\langle \phi_c , V | \int_V d^d x\ \phi (x) / V |\phi_c , V \rangle = \Phi_c .$$
Let $N = V/L_c^d$. One can show, using the
assumption of a finite correlation length, that these states have
the following properties \vskip.3in

\noindent 1.The quantum dynamics of the variable $\Phi_c$ is
amenable to the semi-classical approximation, with expansion
parameter $\propto 1/N$.\vskip.2in

\noindent 2.The matrix elements of local operators between states
with different values of $\Phi_c$ satisfy
$$ \langle \Phi_c , V | \phi_1 (x_1)\ldots \phi_n (x_n) |
\Phi_c^{\prime} , V \rangle \sim e^{-c N} ,
$$ where $n$ is kept finite as $N \rightarrow \infty$.\vskip.2in

\noindent 3.The interference terms in superpositions between states
with different values of $\Phi_c$ remain small for times of order
$N$. This follows from the previous remark and the fact that the
Hamiltonian is an integral of local operators. This remark is proved
by thinking about which term in the t-expansion of $e^{-iHt}$ first
links together the different superposition sectors with an amplitude
of order $1$.   One needs terms of order $N$ in order to flip a
macroscopic number of local clusters from one macro-state to
another. This term is negligible until $ t \sim N$ in units of the
correlation length. For many systems, there is a technical problem
in this argument, because the Hamiltonian is unbounded, but it is
intuitively clear that a cutoff at high energy should not affect the
infrared considerations here.

\noindent 4. However, there is another important phenomenon
occurring, on a much shorter time scale. The microscopic degrees of
freedom are evolving according to the microscopic Hamiltonian,
perturbed by the time dependent term due to the motion of the
collective coordinate. In a typical situation, the macroscopic
object is not in its quantum ground state, but rather in some
micro-state that is a superposition of many eigenstates from an
energy band where the density of states is, according to quantum
field theory, of order $e^N$\footnote{In reality, the system is
unavoidably coupled to an environment, and is not in a pure state.
If nothing else, soft photon emission will create such an
environment. However, since our point is that the macroscopic system
decoheres itself, we can neglect the (perhaps numerically more
important) environmental decoherence.}. The evolution in this
subspace of states is qualitatively like that of a random
Hamiltonian in a Hilbert space of this dimension. It leads to
thermalization and loss of quantum coherence, through rapid changes
of relative phase\cite{berrysrednicki}. Initial quantum correlations
will reassert themselves only once a recurrence time, and the
shortest recurrence time is $o(e^N)$.

In the language of the previous section, {\it averages of local
fields over distances large compared to the correlation length are
good pointer observables, whenever the system is in a typical state
chosen from an ensemble where the density of states is $o(e^N )$}.
It is only when a macro-system is in its ground state, and the
motion of the collective coordinates is adiabatic, relative to the
gap between the ground state and the region of the spectrum with
exponential density of states, that conventional estimates of power
law (in $N$) time scales for seeing quantum coherence are valid.

Typical field theories describing systems in the real world contain
hydrodynamic modes like phonons, with very low energies, and one
would have to consider frequencies of collective coordinate motion
lower than these hydrodynamic energy scales in order to observe
quantum coherence for the macroscopic observables over power law
time scales.  Certainly the motion of the moon is not in such an
adiabatic regime.

To define an actual apparatus, we have to assume that the quantum
field theory admits bound states of arbitrarily large size.
Typically this might require us to add chemical potential terms to
the Hamiltonian and insist on macroscopically large expectation
values for some conserved charge.  The canonical example would be a
large, but finite, volume drop of nuclear matter in QCD.  We can
repeat the discussion above for averages over sub-volumes of the
droplet.

Of course, in the real world, the assumption of a microscopically
small correlation length is not valid, because of electromagnetic
and gravitational forces. Indeed, most real measuring devices use
these long range forces, both to stabilize the bound state and for
the operation of the machine itself. I do not know how to provide a
mathematical proof, but I am confident that the properties described
above survive without qualitative modification\footnote{In the
intuitive physics sense, not that of mathematical rigor.}. This is
probably because all the long range quantum correlations are
summarized by the classical electromagnetic and gravitational
interactions between parts of the system\footnote{ Recall that the
Coulomb and Newtonian forces between localized sources are described
in quantum field theory as quantum phase correlations in the wave
function for the multi-source system.} . It would be desirable to
have a better understanding of the modification of the arguments
given here, that is necessary to incorporate the effects of
electromagnetism and (perturbative) gravitation. One may also
conclude from this discussion that a system at a quantum critical
point, which has long range correlations not attributable to
electromagnetism or gravitation, would make a poor measuring device,
and might be the best candidate for seeing quantum interference
between ``macroscopic objects".  Of course, such conformally
invariant systems do not have large bound states which could serve
as candidate ``macroscopic objects".

Despite the mention of gravitation in the previous paragraph, the
above remarks do not apply to regimes in which the correct theory of
quantum gravity is necessary for a correct description of nature. We
are far from a complete understanding of a quantum theory of
gravity, but this author believes that it is definitely not a
quantum field theory. In a previous version of this article
\cite{classical1} I gave a brief description of my ideas about the
quantum theory of gravitation.  I believe that it gets in the way of
the rest of the discussion, and I will omit all but the conclusions.

In my opinion, the correct quantum theory of gravity has two sorts
of excitations, something resembling conventional particles, and
black holes. A given region of space-time supports only a finite
amount of information, and a generic state of that region is a black
hole. Black holes have very few macroscopic moving parts, and do not
make good measuring devices.  Low entropy states in the region can
be described in terms of particles with local interactions, and are
approximable for many purposes by local field theory. I have
explained how the general principles of field theory lead to an
understanding of approximately classical measuring devices.

The exponential approach to the classical limit allows us to
understand why these conclusions will not be changed in the quantum
theory of gravity. Systems describable by local field theory over a
mere $10^3 - 10^4$ correlation volumes already have collective
variables so classical that their quantum correlations are
unmeasurable. The fact that there exist energy scales orders of
magnitude below the Planck scale, when combined with these
observations, show us that practically classical systems can be
constructed without the danger of forming black holes.

On the other hand, these same considerations show us that {\it
exactly classical} observables in a quantum theory of gravity must
be associated with infinite boundaries of space-time.  This
observation is confirmed by existing string theory models, and has
profound implications for the construction of a quantum theory of
gravity compatible with the world we find ourselves in.
\section{Proposed experiments}

\subsection{Schr\"{O}dinger's drum}

I was motivated to rewrite this article for publication, by a number
of papers, which propose experiments to observe quantum correlations
for the observables of a mesoscopic system\cite{schrdrum}. In its
simplest form the system consists of two dielectric membranes,
suspended in a laser cavity. By tuning the laser frequencies, it is
claimed that one can cool the motion of the translational collective
coordinates of the two membranes, which are coupled through the
laser modes, down to their ``steady state ground state".  The ground
state can be engineered to be a superposition of two different
relative positions for the membranes.  The system has been dubbed
{\it Schr\"{o}dinger's Drum}\cite{schrdrum}. Although state of the
art experiments can not yet reach the ground state splitting, it is
plausible that it can be reached in the near future.

The membranes are about one millimeter square and 50 nano-meters
thick. Typical phonon energies are thus of order $10^{-4}$
eV\footnote{Here I refer to the phonons of internal sound waves on
each membrane, rather than the phonons associated with the relative
motion of the two membranes.}.   In the analysis of the proposed
experiments \cite{p} it is argued that the collective coordinates of
the two membranes in a laser cavity has a pair of classically
degenerate ground states, which are split by $10^{-10}$ eV.  It is
then argued that by tuning the laser frequencies, one can cool the
collective coordinate system down to temperatures below this
splitting.  The true ground state is a superposition of two
classical values for the collective coordinates, and it is claimed
that one can observe the entanglement of these two states.

At first glance, one might assume that the extremely low temperature
of the collective coordinates means that the considerations of our
analysis are irrelevant.  However, on closer scrutiny it becomes
apparent that the whole process of laser cooling, depends crucially
on the coupling between the collective coordinates and a source of
``mechanical noise".  The latter is treated as a large thermal
system with a temperature (in the theoretical analysis $10^{-7} -
10^{-6}$ eV)\footnote{I suspect that when applied to actual
experiments with mm. size membranes, the source of this noise is
excitation of sound waves on the membrane, and the temperature is
even higher.}.  The analysis of \cite{p} uses a quantum Langevin
equation to describe the way in which energy is drained from the
collective coordinates into the reservoir of mechanical noise. This
might be\footnote{The book \cite{qn} referred to in \cite{p}
suggests that the quantum Langevin treatment is adequate only when
the noise reservoirs are collections of oscillators with linear
couplings to the collective coordinates. It is not clear to me that
this is the case.  The collective coordinates are really zero wave
number phonons of the individual membranes, and I would have guessed
that they are coupled non-linearly to the shorter wavelength modes.}
perfectly adequate for showing that the temperature of the
collective coordinates can indeed be lowered, but it does not give
an adequate account of quantum phase coherence.

The very fact that the source of mechanical noise can be modeled as
a system obeying the laws of statistical mechanics, implies that the
collective coordinate is coupled to a large number of other degrees
of freedom, in a regime where the density of states of these degrees
of freedom is exponentially large.  Our analysis applies, and there
should be no phase coherence between superselection sectors of the
noise bath.  If we take the correlation length in the membranes to
be $100$ nano-meters, then $e^N \sim e^{10^{12}}$, and there is no
hope of seeing quantum coherence.  {\it The failure to see quantum
correlations in these experiments is not an indication that quantum
mechanics breaks down for macro-systems, but simply a failure to
understand that the approximate two state system of the membrane
collective coordinates, suffers decoherence due to its coupling to
the system which cools it down to the quantum energy regime.}

Indeed, the collective coordinates are coupled to the the system
that provides the mechanical noise. The very fact that it is
permissible to describe this system by statistical mechanics shows
that it has a huge reservoir of states through which it is cycling
on a microscopic time scale.  The coupling of the collective
coordinates to these states, which is necessary for the cooling
process, also washes out phase correlations between different
classical states of the collective modes.

\subsection{Josephson's flux}

By contrast, an older experiment\cite{friedman} seems to illustrate
the fact that when the microscopic degrees of a macro-system can be
kept in their ground state, the standard analysis of the quantum
mechanics of collective coordinates is correct.   This experiment
consists of a Josephson junction, with a flux condensate composed of
$o(10^9 )$ Cooper pairs. By appropriate tuning, one can bring the
system to a state where there is resonant tunneling between a
degenerate pair of quantum levels of the Landau-Ginzburg order
parameter.  The author's argue, correctly I believe, that because of
the superconducting gap, and because their external magnetic probes
couple directly to the order parameter, they can keep the system in
its quantum ground state.  They verify the level repulsion of a two
state quantum system, when two classically degenerate states are
connected by a tunneling transition.   This experiment truly
achieves a quantum superposition of macro-states.   The recurrence
time scale $e^{10^9}$ of a typical state of this many microscopic
constituents is irrelevant to the analysis of this experiment.

There is a hint here of what is necessary in order to approach
macroscopic superpositions, and it echoes an insight that has
already appeared in the quantum computing literature.
Kitaev\cite{kitaev} has emphasized that {\it topological order
parameters} may be essential to the construction of a practical
quantum computer.  Quantum Hall systems and superconductors have
such order parameters.  In the rather abstract language of quantum
field theory, what would appear to be necessary is a system whose
low energy dynamics is described by a ${\it topological field
theory}$.  In plain terms, this is a system whose localized
excitations are separated from excitations of a selected set of
topological variables by an energy gap.  From a practical point of
view, what we need is a system in which this gap is large enough so
that one can carry out experiments, which do not excite states above
the gap.

Macroscopic systems will generically have phonon excitations, with
energies that scale like the inverse of the largest length scale in
the macroscopic body.  However, as the case of the Josephson
junction shows, it may be possible to devise probes of the system,
which couple directly to the topological order parameters, without
exciting mechanical oscillations. If that is the case, we are in a
regime which is properly analyzed by the conventional collective
coordinate quantum mechanics.   For appropriately mesoscopic
systems, and with sufficient protection against decoherence by
coupling to a larger environment, we can achieve quantum coherence
for appropriate macroscopic order parameters.

This does not change the main burden of this article, which is that
for typical macroscopic objects, quantum coherence is {\it
superexponentially} unlikely, and cannot be observed over any
experimentally realizable time scale.  It does however, confirm the
insight of Kitaev, that there may be a topological route to
practical quantum computation.

\section{\bf Conclusions}

I suspect the material in this paper is well understood by many
other physicists, including most of those who have worked on the
environmental decoherence approach to quantum measurement. If there
is anything at all new in what I have written here about quantum
measurement, it lies in the statement that a macroscopic apparatus
of modest size serves as its own ``environment" for the purpose of
environmental decoherence. In normal laboratory circumstances, the
apparatus interacts with a much larger environment and the huge
recurrence and coherence times become even larger. Nonetheless,
there is no reason to suppose that a modestly macroscopic apparatus,
surrounded by a huge region of vacuum, with the latter protected
from external penetrating radiation by thousands of meters of lead,
would behave differently over actual experimental time scales, than
an identical piece of machinery in the laboratory.

The exception to this kind of self-decoherence that we have
identified, seems to involve topological variables of systems like
superconductors and quantum Hall materials.  These are systems with
an interesting finite dimensional Hilbert space of quasi-degenerate
ground states, separated from the rest of the spectrum by a
substantial gap.  In addition, one must have probes which can couple
directly to the topological variables, without exciting low energy
phonon degrees of freedom (which are present in any macroscopic
object).  For such systems, one might expect to be able to create
robust superpositions of states of collective variables of
macroscopic systems.  Kitaev has argued that these may be the key to
quantum computing.

The essential point in this paper is that the corrections to the
classical behavior of macroscopic systems are exponential in the
size of the system in microscopic units. This puts observable
quantum behavior of these systems in the realm of recurrence
phenomenon, essentially a realm of science fiction rather than of
real experimental science. When a prediction of a scientific theory
can only be verified by experiments done over times
super-exponentially longer than the measured age of the universe,
one should not be surprised if that prediction is counter-intuitive
or ``defies ordinary logic".

Quantum mechanics does make predictions for macro-systems which are
different than those of deterministic classical physics. Any time a
macro-system is put into correlation with a microscopic variable -
and this is the essence of the measurement process - its behavior
becomes unpredictable. However, these predictions are
indistinguishable from those of classical statistical mechanics,
with a probability distribution for initial conditions derived from
the quantum mechanics of the micro-system. It is only if we try to
interpret this in terms of a classical model of the micro-system
that we realize something truly strange is going on. The predictions
of quantum mechanics {\it for micro-systems} {\it are} strange, and
defy the ordinary rules of logic. But they do obey a perfectly
consistent set of axioms of their own, and we have no real right to
expect the world beyond the direct ken of our senses, which had no
direct effect on the evolution of our brains, to "make sense" in
terms of the rules which were evolved to help us survive in a world
of macroscopic objects.

Many physicists, with full understanding of all these issues, will
still share Einstein's unease with an intrinsically probabilistic
theory of nature. Probability is, especially when applied to
non-reproducible phenomena like the universe as a whole, a theory of
guessing, and implicitly posits a mind, which is doing the guessing.
Yet all of modern science seems to point in the direction of mind
and consciousness being an emergent phenomenon; a property of large
complex systems rather than of the fundamental microscopic laws. The
frequentist approach to probability does not really solve this
problem. Its precise predictions are only for fictional infinite
ensembles of experiments. If, after the millionth toss of a
supposedly fair coin has shown us a million heads, and we ask the
frequentist if we're being cheated, all he can answer is
``probably". Neither can he give us any better than even odds that
the next coin will come up tails if the coin toss is truly unbiased.

I have no real answer to this unease, other than ``That's life. Get
over it."  For me the beautiful way in which linear algebra
generates a new kind of probability theory, even if we choose to
ignore it and declare it illogical\footnote{One can easily imagine
an alternate universe, in which a gifted mathematician discovered
the non-commutative probability theory of quantum mechanics, and
speculated that it might have some application to real measurements,
long before experimental science discovered quantum mechanics.}, is
some solace for being faced with a question to which, perhaps, my
intrinsic makeup prevents me from getting an intuitively satisfying
answer. On the other hand, I believe that discomfort with an
intrinsically probabilistic formulation of fundamental laws is the
only ``mystery" of quantum mechanics.  If someone told me that the
fundamental theory of the world was classical mechanics, with a
fixed initial probability distribution, I would feel equally
uncomfortable. The fact that the laws of probability for
micro-systems don't obey our macroscopic ``logic" points only to
facts about the forces driving the evolution of our brains. If we
had needed an intuitive understanding of quantum mechanics to obtain
an adaptive advantage over frogs, we, or some other organism, would
have developed it. Perhaps we can breed humans who have such an
intuitive understanding by making the right to reproduce contingent
upon obtaining tenure at a physics department. Verifying the truth
of this conjecture would take a long time, but much less than time
than it would take to observe quantum correlations in a
superposition of macro-states.

\vskip.4in \section{\bf Acknowledgments}

I would like to thank Michael Nauenberg, Anthony Aguirre, Michael
Dine, Jim Hartle, W. Zurek, Bruce Rosenblum, and especially Lenny
Susskind, for important discussions about this thorny topic. This
research was supported in part by DOE grant number
DE-FG03-92ER40689. \vfill\eject

\end{document}